\def\gsim{\mathrel{\scriptstyle{\buildrel > \over \sim}}}

\magnification 1200
\baselineskip=17pt

\centerline{\bf SKYRMION LIQUID PHASE OF THE QUANTUM}
\bigskip
\centerline{\bf FERROMAGNET IN TWO DIMENSIONS}
\vskip 50pt
\centerline{J. P. Rodriguez}
\smallskip
\centerline{\it Instituto de Ciencia de Materiales (CSIC),
Universidad Autonoma de Madrid,}

\centerline {\it Cantoblanco, 28049 Madrid, Spain.}
\centerline{\it and}
\centerline{{\it Dept. of Physics and Astronomy,
California State University,
Los Angeles, CA 90032.}\footnote*{Permanent address.}}

\vskip 30pt
\centerline  {\bf  Abstract}
\vskip 8pt\noindent
The two-dimensional quantum ferromagnet filled with a liquid
of skyrmions is studied theoretically in the context of the quantum
Hall effect near electronic filling factor $\nu = 1$.
A cross-over between the classical ferromagnetic phase at $\nu =1$
and a classical paramagnetic phase at $\nu \neq 1$ is obtained,
which is consistent with recent Knight-shift measurements.
A new collective mode associated with the skyrmion liquid that is of
the cyclotron type is also identified.
\bigskip
\noindent
PACS Indices: 73.40.Hm, 75.10.Jm, 75.40.Cx
\vfill\eject

It has recently been proposed that skyrmion spin textures
are tied to excess electronic charge with respect to 
the quantum Hall state at unit filling,$^{1,2}$  the spin arrangement
of which is known to be a two-dimensional (2D) ferromagnet.
Crudely speaking, a skyrmion is simply a center of reversed
spins within the 2D ferromagnet.  The  2D environment is crucial,
however, to insure the topological stability of these objects.$^{3-5}$
Experimental evidence in support of the former proposal
has come from nuclear magnetic resonance  (NMR) studies
of the quantum Hall state, where a rapid degradation in the total
spin polarization as measured from the Knight shift
has been observed as a function of the deviation from
unit filling.$^6$  Optical and magnetotransport
measurements have reached similar
conclusions.$^7$   Furthermore, zero-temperature Hartree-Fock calculations
based on a crystalline background of skyrmions have obtained
quantitative agreement with the experimentally observed dependence
of the spin polarization with the electronic filling factor
at low temperature.$^8$  It is natural to wonder, however,
whether a skyrmion liquid$^9$ (or gas$^{10}$) state is more likely
to occur in such quantum Hall states at non-zero temperature. 
In fact, recent specific heat measurements,$^{11}$ as well as theoretical
calculations,$^{12}$ indicate that the skyrmion crystal melts at a 
low temperature with respect to the Zeeman   energy splitting in
the quantum Hall system.

Motivated by the magnetic phenomena displayed by the quantum
Hall state near unit filling, we shall study here the 2D quantum
ferromagnet in the presence of a homogeneous net density
of skyrmions.  
The residual Coulomb interaction inherited by the
skyrmions in the context of the quantum Hall effect 
will be neglected, however, which means that
the liquid nature of the system is exaggerated.$^{13}$
The theoretical discussion begins with a meanfield analysis
of the CP$^1$ model for the quantum ferromagnet$^{14,15}$  in two
dimensions at constant skyrmion density.$^{10}$  The density
of skyrmions in the ferromagnet acts as a continuous dial
between the following regimes
(see Fig. 1): a skyrmion-poor 2D ferromagnet at extremely low density; a
skyrmion-rich classical paramagnet at higher densities 
that is characterized
by a ferromagnetic correlation length limited by the separation
between skyrmions; and finally an  ideal quantum paramagnet at 
extremely  high
density.  Note that no phase transitions 
separate these regions.  The central result of the mean-field analysis is
the magnetization (see Fig. 2), which is
found to give a fair comparison with available Knight-shift
data on the quantum Hall state.  A new collective mode
connected with the skyrmion liquid is also found, however,
once  fluctuations with respect to the meanfield saddle point
are accounted for.
It can be
understood in physical terms as a collective precession of the
boundary spins,
which is equivalent to 
the orbit
of a skyrmion around the edge of the ferromagnet.  As expected
then, no Goldstone
mode appears in the paramagnetic skyrmion liquid.

Our task then is to determine the thermodynamic and dynamic properties
of the 2D ferromagnet in the presence of both a homogeneous density
of skyrmions and of an external magnetic field.  As a means to
this end, we introduce the partition function
$$Z = \int {\cal D} a_0^{\prime} 
{\cal D}\vec a^{\prime} {\cal D} z {\cal D} \bar z\,
{\rm exp}(-S/\hbar) \eqno (1)$$
for the corresponding CP$^1$ model,$^{14,15}$
whose action $S$ in imaginary time
$\tau = it$  reads
$$\eqalignno{ S = \int_0^{\hbar/k_B T} d\tau\int d^2 r\{\hbar
M_0[\bar z (\partial_{\tau}-a_0^{\prime})z +
a_0^{\prime} +  {\rm c.c.}]&
+ 2\rho_s|(\vec\nabla -i\vec a^{\prime})z|^2 +\cr
&+ M_0 E_Z(|z_-|^2 - |z_+|^2)\}.&(2)}$$
Here $M_0 = s a^{-2}$ is the fully polarized magnetization
for a density $a^{-2}$ of spin $s$ moments, $\rho_s = s^2 J$ is the
spin stiffness of the ferromagnet, and $E_Z =  \mu g H$
is the Zeeman energy splitting.  The complex doublet field
$z = (z_-, z_+)$ is connected to the normalized magnetization
by $\vec m = \bar z\vec\sigma z$, where $\vec\sigma$ denotes the Pauli
matrices with the $z$-axis directed along the external magnetic field.
Integration (1) over the scalar potential $a_0^{\prime}$ 
and over the vector potential
$\vec a^{\prime}$ in the Coulomb
gauge ($\vec\nabla\cdot\vec a^{\prime} = 0$) enforces the 
respective constraints
$|z|^2 = 1$ and $\vec a^{\prime} = i(\vec\nabla\bar z) z$ 
characteristic of the CP$^1$ model for the
2D ferromagnet.$^{4,5,10}$  The 3-vector
$J_S^{\mu} = (n_S, \vec J_S)$ for the local skyrmion
density, $n_S$, and local current density, $\vec J_S$, is also
related to these potential fields by the manifestly gauge-invariant
formula
$J_S^{\mu} = (2\pi)^{-1} \epsilon^{\mu\nu\gamma}
\partial_{\nu} a_{\gamma}^{\prime}$,
 where $a_{\mu}^{\prime} = (a_0^{\prime}, \vec a^{\prime})$ and 
$\partial_{\mu} = (\partial_t, \vec\nabla)$.
(Repeated Greek indices are hereafter summed over.)
This identity  can be understood by noticing that
$n_S = (\partial_x a_y^{\prime} - \partial_y a_x^{\prime})/2\pi$
gives the local skyrmion density of the classical
ferromagnet,$^{10}$  while the 3-vector $J_S^{\mu}$
formally satisfies the continuity equation,
$\dot n_S +\vec\nabla\cdot\vec J_S = 0$.$^{3-5}$

In the meanfield approximation we presume constant
potentials $a_{\mu}^{\prime} = (a_0^{\prime}, 0, b_S x)$ 
and ignore fluctuations.  Here, 
the fictitious
magnetic field $b_S$ is tied to the average skyrmion density by
$n_S = b_S/2\pi$, which is a conserved quantity. 
Integrating
(1) over the $z$-fields and then minimizing the resulting
free energy with respect to $a_0^{\prime}$, we obtain the
averaged constraint
$1 = \langle\bar z_- z_-\rangle + \langle\bar z_+ z_+\rangle$, where
$\langle\bar z_{\pm} z_{\pm}\rangle
= {1\over 2}(n_S/M_0)\sum_{n=0}^{\infty}
[{\rm exp}(E_n^{\pm}/k_BT) - 1]^{-1}$ is 
the trace of the
propagator for the
$z_{\pm}$ quantum, with Landau-type energy levels
$E_n^{\pm} = (a^2/s)(\xi^{-2} + 2b_S n)\rho_s \mp {1\over 2}E_Z$.
This sets
the unknown $\xi^{-2}\propto a_0^{\prime}$,  which 
then allows us to compute the normalized magnetization
along the z-direction, 
$m_z = \langle\bar z_+ z_+\rangle - \langle\bar z_- z_-\rangle$,
as a function of temperature.  The Landau levels in the present
mean-field theory have an energy
splitting of $\hbar\omega_c^{\prime} = (n_S/M_0) E_S$,
where $E_S = 4\pi\rho_s$ is the classical energy cost of a single
skyrmion.$^{3-5}$  Let us now analyze these mean-field equations along
the skyrmion-density axis.
In the skyrmion-poor limit 
$\hbar\omega_c^{\prime}\ll k_B T$, for example,
the sum over Landau levels may be converted into an  energy integral.
We thereby recover known results
for the pure quantum 2D ferromagnet.$^{15}$ 
In particular, at
low fields, $E_Z < E_S$, this pure ferromagnet has three distinct
phases (see Fig. 1): ({\it i}) a low-temperature ($k_BT\ll E_Z$)
quantum activated regime
characterized by the saturated magnetization
$m_z\cong 1 - (k_B T/E_S)e^{-E_Z/k_BT}$; 
({\it ii}) a high-temperature ($E_S\ll k_BT$) quantum-critical
regime that follows the Curie law $m_z\cong E_Z/2k_BT$;
({\it iii}) and a classical regime at intermediate temperatures
($E_Z\ll k_BT\ll E_S$) 
that interpolates
smoothly between the previous two behaviors.$^{15}$
The ferromagnetic correlation length in the latter
classical ferromagnetic phase notably diverges exponentially
as $\xi_{\rm FM}\propto e^{E_S/2k_BT}$.  No critical phenomena separate
the phases mentioned above.

On the other hand, in the opposing skyrmion-rich limit,
$\hbar\omega_c^{\prime} \gg k_B T$, it becomes sufficient to
retain only the lowest Landau-level term ($n=0$) in mean-field
sums.
The spin-spin correlation function is in general proportional
to that of the $z$-quanta squared, which in the present case is given
by$^{10}$ $|\langle z(\vec r) \bar z(\vec r\,^{\prime})\rangle|^2
= {\rm exp}(-{1\over 2}|\vec r - \vec r\,^{\prime}|^2 b_S)$. 
This means that the ferromagnetic correlation length in the
skyrmion-rich regime is limited by the separation between neighboring
skyrmions;  i.e., $\xi_{\rm FM}\sim  n_S^{-1/2}$.
Upon further  analysis of the low-field case $E_Z < E_S$
in this limit, the following regimes can be identified (see Fig. 1):
({\it i})  a quantum-activated regime at low temperature $k_BT\ll E_Z$
and at low skyrmion density $n_S/M_0\ll 1$, with a saturated
magnetization $m_z \cong 1 - (n_S/M_0)e^{-E_Z/k_BT}$;
({\it ii}) an ideal quantum paramagnet at extremely high
skyrmion densities $n_S/M_0\gg 1$, with a corresponding 
magnetization
$$m_z^{\rm qp} = {\rm tanh}(E_Z/2k_BT); \eqno (3)$$
({\it iii}) and a classical paramagnet in the remainder of the
phase diagram, $n_S/M_0\ll 1$ and $E_Z \ll k_BT \ll E_S$,
 with a normalized magnetization
$$m_z^{\rm cp} = [1+ (T/T_*)^2]^{1/2} - T/T_*, 
\eqno (4)$$
where $k_BT_* = (M_0/n_S) E_Z$ is the crossover scale
(see ref. 10).  Notice that the magnetization of
the classical paramagnet follows a Curie law, 
$m_z\cong T_*/2T$, for temperatures $T\gg T_*$.
Again, no critical phenomena separate
the various regimes.

We shall now obtain  the structure  of the collective modes in the
skyrmion liquid by allowing for fluctuations with respect the
above meanfield saddlepoint.  In particular, dynamical fluctuations
in the fictitious gauge field, $\delta a_{\mu}^{\prime}$, which 
are equivalent to fluctuations in the skyrmion density and current,
contribute the standard second-order term
$$S_2 = {i\over{2\hbar}}\int {d\omega\over{2\pi}}\int {d^2k\over{(2\pi)^2}}
[\Pi_{\mu\nu}^+(\vec k, \omega) + \Pi_{\mu\nu}^-(\vec k,\omega)]
\delta a_{\mu}^{\prime} (\vec k, \omega)
\delta a_{\nu}^{\prime} (-\vec k, -\omega)
\eqno (5)$$
to the action (2), where
$$\eqalignno{
\Pi_{\mu\nu}^{\pm}(\vec k,\omega) = &
V^{-1}\sum_{n, q} 2 n_B(E_n^{\pm})
(\hat x_{\mu}\hat x_{\nu} + \hat y_{\mu}\hat y_{\nu}) \cr
&-V^{-1}\sum_{n\neq n^{\prime}, q}
{n_B(E_n^{\pm}) - n_B(E_{n^{\prime}}^{\pm})\over
{\hbar\omega - E_n^{\pm} + E_{n^{\prime}}^{\pm}}}
\langle n,q|j_{\mu} (\vec k)|n^{\prime}, q\rangle
\langle n^{\prime}, q|j_{\nu}(-\vec k)|n, q\rangle 
& (6)\cr}$$
is the Kubo-Lindhard formula  corresponding to  the $z_{\pm}$ quantum
for wavenumber $\vec k$ chosen to be parallel to the $x$-axis.
Here $n_B(E_n^{\pm})$ denotes the Bose distribution function,
$j_0(\vec k) = e^{ikx}$ and 
$\vec j (\vec k) = 2 (a^2/s) \rho_s e^{ikx/2} (i\vec\nabla +\vec a^{\prime})
e^{ikx/2}$ are the respective density and current operators,
and $\langle\vec r|n, q\rangle = \langle x^{\prime}|n\rangle
e^{iqy}$ are the eigenstates
for Landau levels in the gauge $\vec a^{\prime} = (0, b_S x)$, where
$|n\rangle$ denote the standard harmonic oscillator states,
with $x^{\prime} = x - b_S^{-1} q$.
Expression (6) can be evaluated explicitly in the long-wavelength
limit, $\vec k\rightarrow 0$.
Taking    the Landau gauge ($\delta a_x^{\prime} = 0$), 
one obtains a   polarizibility
tensor of the form $\Pi_{\mu\nu}^+ +  \Pi_{\mu\nu}^- =
\epsilon (\omega) \pi_{\mu\nu}$ at  frequencies
$\omega$ near $\omega_c^{\prime}$, with a
dielectric function
$\epsilon(\omega) = 4\rho_s/(\omega_c^{\prime 2} - \omega^2)$,
and with components $\pi_{yy} = c_1^2 k^2 - \omega^2$,
$\pi_{0y}  = ik\omega_c^{\prime}$, $\pi_{y0} = -\pi_{0y}$,
and $\pi_{00} = - k^2$.  Here, $c_1 = (3/8\pi)^{1/2}
(n_S^{1/2}/M_0) (E_S/\hbar)$ is the dispersion
velocity in the skyrmion-rich limit.  The collective mode, by definition,
is the null eigenstate of the previous polarizibility
tensor.  It appears at frequencies
$\omega_c^{\prime}(k) = (\omega_c^{\prime 2} + c_1^2 k^2)^{1/2}$,
with  skyrmion
density-current components 
$(\delta n_S, \vec J_S)\propto [k, (\hat x + i\hat y)\omega_c^{\prime}]$
of the cyclotron type.
Notice that the residue  
$1/\epsilon[\omega_c^{\prime}(k)]$ of the pole
in the response function,
$\langle \delta a_{\mu}^{\prime} 
\delta a_{\nu}^{\prime}
\rangle$, 
corresponding to
this collective mode varies as $c_1^2 k^2$, which vanishes in 
the long-wavelength limit.

The above result for the
cyclotron-type collective mode connected with the skyrmion
liquid is notably independent
of temperature.  This  suggests that it
has a mechanical basis, which we now
outline.  Consider
the edge of the ferromagnet filled with skyrmions.   A    density
$n_S$ of skyrmions in the bulk will then produce    a
normalized  edge magnetization,
$\delta\vec m\sim (n_S a^2) (\hat e\times\hat z)$, where $\hat e$
and $\hat z$ are respectively the unit vectors along the edge
and  along the bulk magnetization.  The presence of such a magnetization
will then induce a precession about it by the spins
at the edge, at an angular
frequency $\omega_c^{\prime} = \rho_s |\delta\vec m|/s\hbar
\sim (n_S/M_0)(\rho_s/\hbar)$.  Yet given that the spin in a
symmetric skyrmion winds around precisely  once along any 
straight line passing through the center of the
skyrmion, then such a precession
is equivalent to the orbit of skyrmions around the
bulk at an angular  frequency $\omega_c^{\prime}$.  Notice that this
argument depends only on the presence of a non-zero density
of skyrmions in bulk, which suggests that the collective
mode is robust with respect to any tendency towards
crystallization$^{8}$ in the skyrmion liquid.
Last, it should be remarked that the same argument may be
employed to demonstrate that two skyrmions must orbit about one 
another.$^{16}$

The proposal that electron (hole) excitations with respect to the
$\nu =1$ quantum Hall state induce
a background  skyrmion spin-texture$^1$ suggests that
the present theory for the
skyrmion-filled ferromagnet 
describes the magnetic properties of quantum Hall liquids in
the vicinity of $\nu  = 1$.
However, the Chern-Simons term
$S_{\rm CS} = (4\pi)^{-1}\int dt\int d^2r\epsilon^{\mu\nu\gamma}
a_{\mu}\partial_{\nu} a_{\gamma}$
must first be added to the ferromagnetic action (2),$^1$
where the statistical gauge field $a_{\mu}$ that transmutes the
original electronic field $\Psi_{\sigma}$ into the bosonic CP$^1$
model field $z_{\sigma}$ is related  by the identity
$a_{\mu}^{\prime} = a_{\mu} + (e/\hbar c) A_{\mu}$ 
to the skyrmionic
gauge field $a_{\mu}^{\prime}$ and to the external electro-magnetic
potentials  $A_{\mu}$.
When this Chern-Simons term is combined with the one (5) that is
dynamically generated by the skyrmion liquid,
$S_2 = ... + (4\pi\nu_S)^{-1}\int dt\int d^2r\epsilon^{\mu\nu\gamma}
a_{\mu}^{\prime}\partial_{\nu} a_{\gamma}^{\prime} + ...$, 
where $\nu_S = n_S a^2$ is 
the skyrmionic filling factor (for $s = 1/2$),
we obtain that 
({\it i})  the Hall resistance is given by
the Ioffe-Larkin composition formula$^{17}$
$\rho_{xy} = (1 + \nu_S)(h/e^2)$, while ({\it ii}) 
the prefactor corresponding to  the dynamically generated
Chern-Simons terms must be replaced by 
$\nu_S^{-1}\rightarrow \nu_S^{-1} +1$ in
the previous discussion concerning the
skyrmionic collective mode.  However, the skyrmionic filling factor
is equal to
$\nu_S = \nu^{-1} - 1$ for electronic filling factors
$\nu < 1$.$^{10}$  This implies that the Hall resistance
is that naively expected for free electrons
in a partially-filled lowest Landau level,  
$\rho_{xy} =  \nu^{-1}(h/e^2)$, 
and that the nature
of the previously discussed collective mode 
remains unchanged for electronic filling factors in the vicinity
of $\nu  = 1$, since $\nu_S\ll 1$ in such
case.  The latter 
implies that the skyrmion/charge hybrids
orbit at a new cyclotron frequency, $\omega_c^{\prime} < \omega_c$.
This  conflicts with the fact that  each electron (hole) must orbit
at the true cyclotron frequency, $\omega_c$.  The  paradox is
resolved as follows:  The skyrmion collective mode is indeed
possible in the long wavelength limit, $k\rightarrow 0$,
in which case it is purely transverse and
shows no charge fluctuations.  As mentioned
before, it corresponds to a precession of the boundary
spins at  angular frequency $\omega_c^{\prime}$ in such case.
The remaining skyrmion cyclotron modes at $k\neq 0$ are
not possible, however, since they must oscillate at the
true cyclotron frequency by charge conservation.
This conflict must then be interpreted as 
a breakdown of  the present mean-field approximation.
We conclude, nevertheless, that the {\it magnetic}
properties of the quantum Hall
state are indeed well described by the ferromagnetic action (2)
in the absence of the Chern-Simons term for low enough skyrmion densities.

To test this  idea, we shall apply our results
to recent Knight-shift measurements$^6$ that probe the
magnetization  of the quantum Hall state
near electronic filling factor $\nu = 1$
and at  magnetic fields corresponding to
a Zeeman splitting of  $E_Z\cong 2\,{\rm K}$.  
First, the present theory predicts a cross-over from classical
ferromagnetic behavior at $\nu = 1$ to classical 
paramagnetic behavior  away from $\nu = 1$
at a skyrmionic filling factor of
$\nu_S^* = s k_B T/E_S$ (see Fig. 1).  The latter regime is
characterized by a magnetization that depends
strongly on the skyrmion density.
Fixing  the temperature at the Zeeman energy, $k_B T = E_Z$,
then yields a cross-over electronic filling factor
of $\nu_* = 0.97$, which is consistent with
experiment.$^6$
Here the theoretical value for the  spin-stiffness expected 
of  an ideal 2D quantum Hall state at $\nu = 1$ is assumed;$^{15}$  i.e.,
$E_S = 40\,{\rm K}$.
In Fig. 2
is also  shown the theoretically predicted magnetization for a  
density of skyrmions,$^{10}$
$n_S/M_0 = 2 (1 - \nu)/\nu \cong 0.3$,  that corresponds
to an electronic filling-factor of $\nu = 0.88$. 
The experimental Knight-shift results, $K_s(T,\nu)$, 
at this filling are also
shown, where the formula $M/M_0 = K_s(T,\nu)/\nu K_s(0,1)$ has
been used to compute the normalized magnetization.  
Although
the experimental results fall  within the bounds of the theory
set by the classical paramagnetic result at $E_S\rightarrow\infty$
and the quantum paramagnetic result (3) at $E_S = 0$,
they lie substantially below the
theoretically expected  curve.
This suggests that the
naive theoretical value
for the ferromagnetic spin stiffness may be too high.  Similar
conclusions were reached by Read and Sachdev$^{15}$ from their
theoretical fits to the magnetization at $\nu =1$.
Indeed, it is known that $\rho_s$ can be reduced by as
much as $70\%$ once corrections due to the
non-zero thickness of typical quantum
Hall effect devices are taken into account.$^1$

Discussions with L. Brey and E. Rezayi are gratefully
acknowledged.  Special thanks go to  S. Sachdev for
pressing the author to examine the Chern-Simons terms.
This work  was supported in
part by  the Spanish Ministry for Education and Culture, as well
as by National
Science Foundation grant DMR-9322427.

\vfill\eject

\centerline{\bf References}
\vskip 16 pt

\item {1.} S.L. Sondhi, A. Karlhede, S.A. Kivelson, and E.H. Rezayi,
Phys, Rev. B {\bf 47}, 16419 (1993).

\item {2.} H.A. Fertig, L. Brey, R. C\^ ot\' e, and
A.H. MacDonald, Phys. Rev. B {\bf 50}, 11018 (1994).
                                                                 
\item {3.} A.A. Belavin and  A.M. Polyakov, Pis'ma
Zh. Eksp. Teor. Fiz. {\bf 22}, 503 (1975) [JETP Lett. {\bf 22}, 245
(1975)].

\item {4.} R. Rajaraman, {\it Solitons and Instantons}
(North-Holland, Amsterdam, 1987).

\item {5.} A.M. Polyakov, 
{\it Gauge Fields and Strings} (Harwood, New York, 1987).

\item {6.} S.E. Barret, G. Dabbagh, L.N. Pfeiffer, K.W. West, and R.
Tycko, Phys. Rev. Lett. {\bf 74}, 5112 (1995); R. Tycko {\it et al.},
Science {\rm 268}, 1460 (1995).

\item {7.}  E.H. Aifer, B.B. Goldberg, and D.A. Broido,
Phys. Rev. Lett. {\bf 76}, 680 (1996);
A. Schmeller, J.P. Eisenstein, L.N. Pfeiffer, and
K.W. West, Phys. Rev. Lett. {\bf 75}, 4290 (1995). 
 
\item {8.} L. Brey, H.A. Fertig, R. C\^ ot\' e and A.H. MacDonald,
Phys. Rev. Lett. {\bf 75}, 2562 (1995).

\item {9.} A.G. Green, I.I. Kogan and A.M. Tsvelik, Phys. Rev. B
{\bf 53}, 6981 (1996).

\item {10.} J.P. Rodriguez, Phys. Rev. B
{\bf 54}, R8345 (1996).

\item {11.} V. Bayot, E. Grivei, S. Melinte, 
M.B. Santos and M. Shayegan, Phys. Rev. Lett. {\bf 76}, 4584 (1996).

\item {12.} A.G. Green, I.I. Kogan and A.M. Tsvelik, 
Phys. Rev. B {\bf 54}, 16838 (1996).

\item {13.} The area of each skyrmion in the classical paramagnetic
phase is known to be $A_S (T) =  (k_B T/E_Z) a^2$ (see 
ref. 10).  At temperatures $k_B T > E_Z$,
it is then much larger than its true size 
at zero temperature, $A_S(0)\gsim a^2$, which results  from the
balance of the Zeeman energy with the residual Coulomb interaction
(see refs. 1 and 8).  Hence, the neglect of the latter 
is valid at such elevated  temperatures, where entropic
forces dominate.

\item {14.} D.P. Arovas and A. Aurbach, Phys. Rev. B {\bf 38}, 316
(1988).

\item {15.} N. Read and S. Sachdev, Phys. Rev. Lett. {\bf 75},
3509  (1995).

\item {16.}  N. Papanicolaou and W.J. Zakrzewski, Phys. Lett. A
{\bf 210}, 328 (1996).

\item {17.} S.-C. Zhang, Int. J. Mod. Phys. B{\bf 6}, 25 (1992).

\vfill\eject
\centerline{\bf Figure Caption}
\vskip 20pt
\item {Fig. 1}  Shown is the schematic phase diagram for the skyrmion
liquid in the low-field 
limit.  All perforated lines designate
crossovers.  In particular, the diagonal dashed line
$(k_BT = \hbar\omega_c^{\prime})$
separates skyrmion-poor and skyrmion-rich regimes.
The classical paramagnet is the only phase 
in which the magnetization 
depends strongly on skyrmion density [see Eq. (4)].

\item {Fig. 2}  The normalized magnetization for the
skyrmion liquid at relatively weak magnetic field is shown.
The squares represent Knight-shift measurements for
the quantum Hall state at electronic
filling-factor $\nu = 0.88$ (see ref. 6,
as well as the discussion in the text).
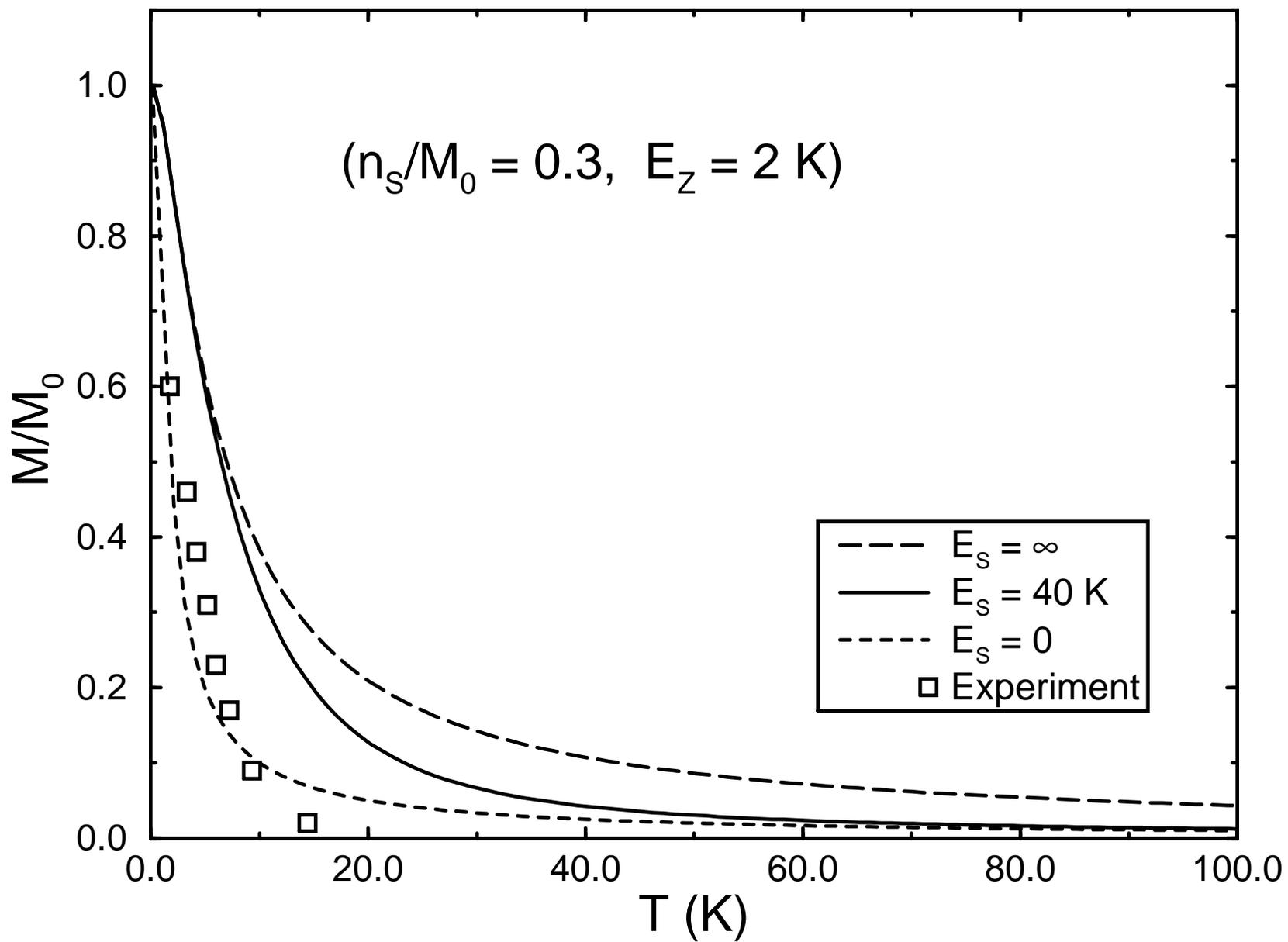
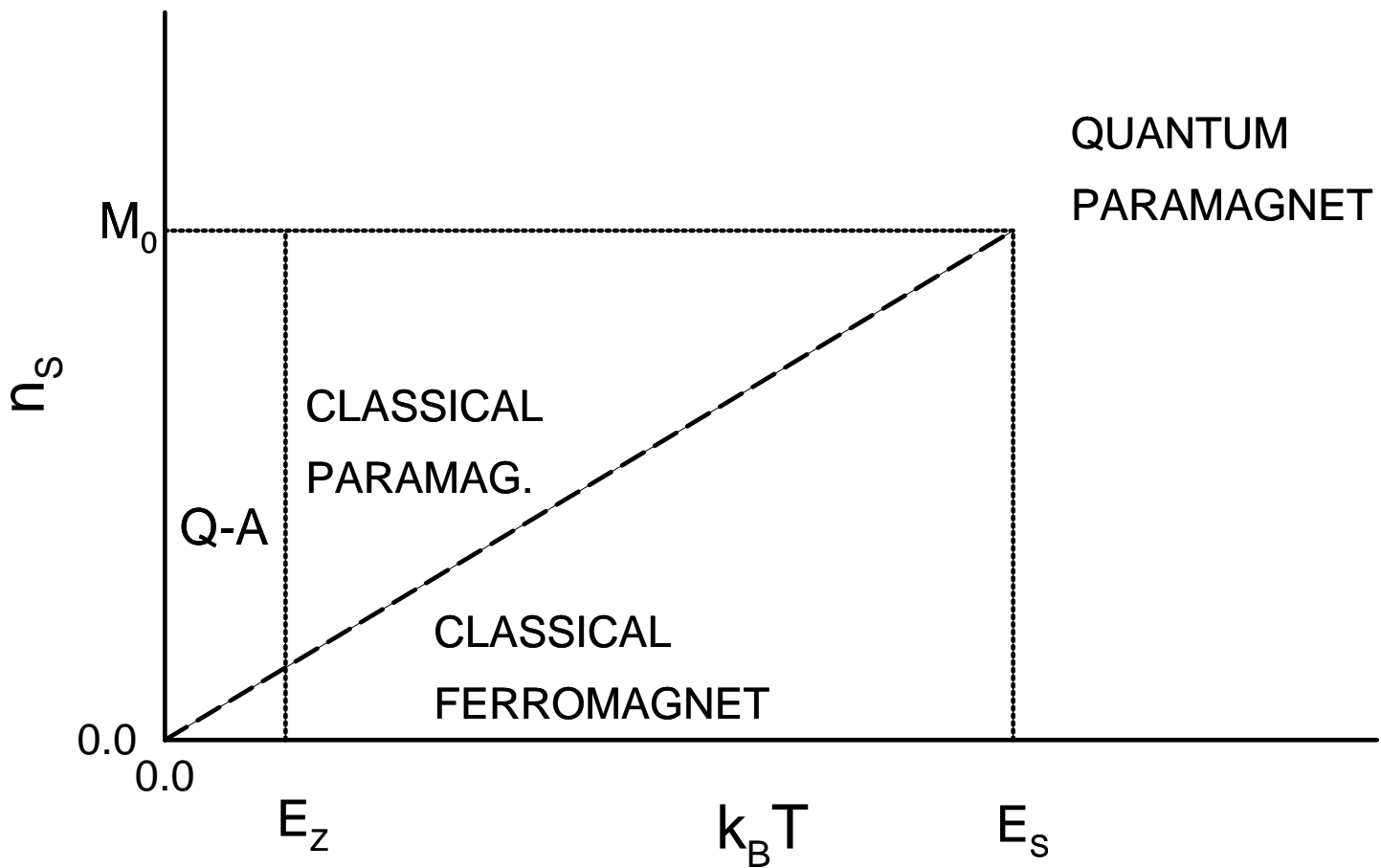
\end